\documentclass[aps,prl,twocolumn,preprintnumbers,superscriptaddress,nofootinbib,10pt]{revtex4-1}

\usepackage{amssymb}
\usepackage{amsmath}
\usepackage{amsfonts}
\usepackage{graphicx}
\usepackage{subfigure}
\usepackage{color}
\usepackage{dcolumn}
\usepackage{hyperref}
\usepackage{comment}

\begin{document}
	
\title{Hydrodynamic and Rayleigh-Plateau instabilities of Q-strings}

\author{Qian Chen}
\email{chenqian192@mails.ucas.ac.cn}
\affiliation{School of Fundamental Physics and Mathematical Sciences, Hangzhou Institute for Advanced Study, University of Chinese Academy of Sciences, Hangzhou, Zhejiang, 310024, China}
\affiliation{Beijing Institute of Mathematical Sciences and Applications, Beijing, 101408, China}

\begin{abstract}
As analogues of compact objects, solitons have attracted significant attention.
We reveal that cylindrical Q-strings exhibit a dynamical instability to perturbations with wavelengths exceeding a threshold $\lambda>\lambda_{c}$. This instability can destroy the invariance in the cylindrical direction, as a generation mechanism for Q-balls, similar to the formation of droplets.
As the interface of Q-strings approaches a thin wall, this long-wavelength instability degenerates into the Rayleigh-Plateau instability with a threshold related only to the geometric radius $\lambda_{c}=2\pi R$.
Such results indicate that Q-strings, like black strings, resemble low-viscosity fluids with surface tension.
\end{abstract}

\maketitle

\begin{figure}
	\includegraphics[width=\linewidth]{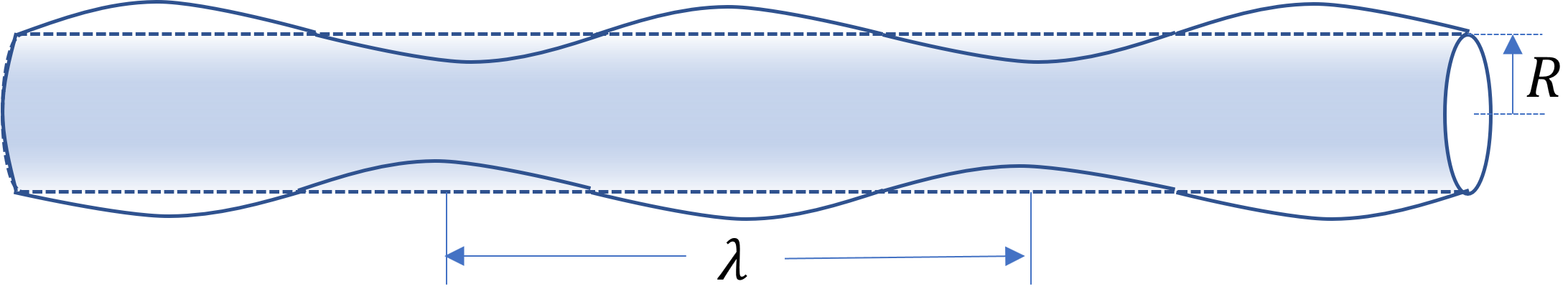}
	\caption{Schematic diagram of Rayleigh-Plateau instability.}
	\label{fig:1}
\end{figure}

{\it Introduction.}---
The formation of droplets, a ubiquitous phenomenon in nature, has fascinated countless theoretical and experimental scientists.
The relevant research has a long history.
As early as 1833 \cite{Savart}, Savart observed fluctuations growing on a jet of water, eventually leading to the formation of droplets.
Such a phenomenon is independent of the circumstance, revealing an intrinsic instability of fluid motion.
After taking into account the effect of surface tension, this instability is demonstrated theoretically by Plateau and Rayleigh \cite{Plateau,Rayleigh1878,Rayleigh1892}, which is explained as the motion of a free surface driven by surface tension.
As shown in Figure \ref{fig:1}, considering a cylinder of fluid with radius $R$, the discovery shows that an arbitrarily small perturbation along the cylindrical surface with a wavelength satisfying $\lambda>2\pi R$ can reduce the surface energy of the system, thereby pushing the configuration to fall exponentially away from equilibrium.
This long-wavelength instability with a geometric threshold is known as Rayleigh-Plateau instability.

This membrane instability is miraculously extended to gravitational systems.
In 1993, Gregory and Laflamme proved that high-dimensional black strings and $p$-branes are linearly dynamically unstable to long-wavelength perturbations \cite{Gregory:1993vy}.
Considering the surface gravity of horizon as the surface tension of fluid membrane, the threshold of the Gregory-Laflamme instability for a black string is consistent with the Rayleigh-Plateau threshold of a hyper-cylindrical fluid flow \cite{Cardoso:2006ks}, demonstrating the similarity between horizons and fluids.
At the nonlinear level, multiple spherical black holes grow on an unstable black string and exhibit self-similarity \cite{Lehner:2010pn}, similar to the formation of droplets in a stream of low-viscosity fluid with the Rayleigh-Plateau instability \cite{Eggers:1997zz}.

For fluids, there is another type of long-wavelength instability related to thermodynamics. 
A typical example is the holographic fluid with thermodynamic instability, which induces a dynamical instability in the sound mode, with a threshold related to the viscosity of the fluid \cite{Janik:2015iry}.
The nonlinear evolution leads to the generation of thermodynamically stable phases \cite{Janik:2017ykj}, where the dynamical behavior can be fully characterized by the {\"M}uller-Israel-Stewart hydrodynamics of a viscous fluid \cite{Attems:2019yqn}.
A similar phenomenon exists in the formation of solitons through the Affleck-Dine mechanism \cite{Affleck:1984fy}.
This type of Affleck-Dine field exhibits both thermodynamic instability \cite{Enqvist:1997si} and spatial instability with long-wavelength \cite{Kusenko:1997si}, resulting in dynamical transitions for thermal phases \cite{Enqvist:1999mv,Kasuya:1999wu,Kasuya:2000wx}.
Such results indicate a potential duality between solitons and fluids, however so far, the relationship between the two remains unknown.

The resulting solitons, especially Q-balls \cite{Coleman:1985ki}, have attracted significant attention with a wide range of applications in modern physics.
They are found to exist in various supersymmetric extensions of the Standard Model \cite{Kusenko:1997zq} and can be copiously produced in the Early Universe, considered as a candidate for dark matter \cite{Kusenko:1997si,Kusenko:2001vu}.
The study of these objects has produced a series of consequences in cosmology, such as the problems of
baryon asymmetry \cite{Enqvist:1997si,Enqvist:1998en}, cosmological phase transitions \cite{Frieman:1988ut} and gravitational waves \cite{Kusenko:2008zm}.
Their extensions in gravity, known as solitonic boson stars \cite{Lynn:1988rb}, serve as models for compact objects.

In the membrane paradigm for black holes, the event horizon is intuitively conceived as a kind of fluid membrane \cite{Thorne:1986iy}.
As analogues of gravitational systems, solitons constructed by matter fields also have interfaces separating matter from the outside world.
An interesting question is whether the interface of solitons behaves similarly to a fluid membrane.
The main result of this letter is to demonstrate for the first time that four-dimensional cylindrical Q-strings exhibit a long-wavelength instability dominated by hydrodynamic modes, serving as a generation mechanism for Q-balls.
As the interface of Q-strings approaches a thin wall, this long-wavelength instability is consistent with the Rayleigh-Plateau instability, revealing the hydrodynamic properties of solitons.



{\it Q-strings.}---We consider the global $U(1)$ symmetric theory involving a self-interacting complex scalar field, described by the following Lagrangian density
\begin{equation}
	\mathcal{L}=-\partial_{\mu}\psi\partial^{\mu}\psi^{*}-V(|\psi|),
\end{equation}
with a commonly used sextic polynomial potential $V=m^{2}|\psi|^{2}-\lambda^{2}|\psi|^{4}+\kappa|\psi|^{6}$,
which can be generated by introducing an additional heavy scalar particle in UV-complete models \cite{Heeck:2022iky}.
Using the dimensionless variables $x:=m x$, $\psi:=\lambda\psi/m$, $\kappa:=m^{2}\kappa/\lambda^{4}$, the potential is simplified to
$V=|\psi|^{2}-|\psi|^{4}+\kappa|\psi|^{6}$.
To ensure that $\psi=0$ is a true vacuum, the single parameter should satisfy $\kappa>1/4$.
In what follows, we will fix $\kappa=0.5$ as an example.

\begin{figure}
	\begin{center}
		\subfigure[]{\includegraphics[width=\linewidth]{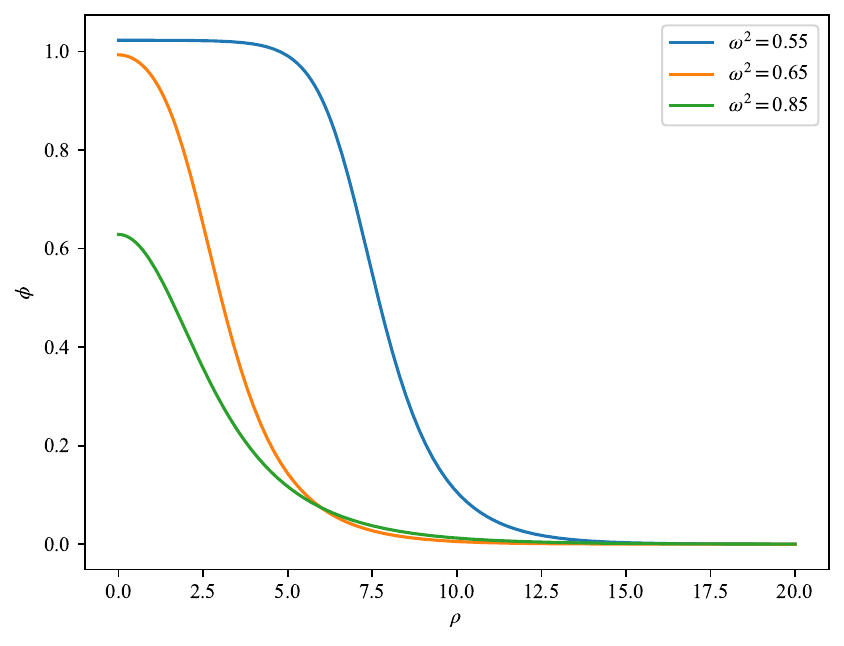}\label{fig:2}}
		\subfigure[]{\includegraphics[width=\linewidth]{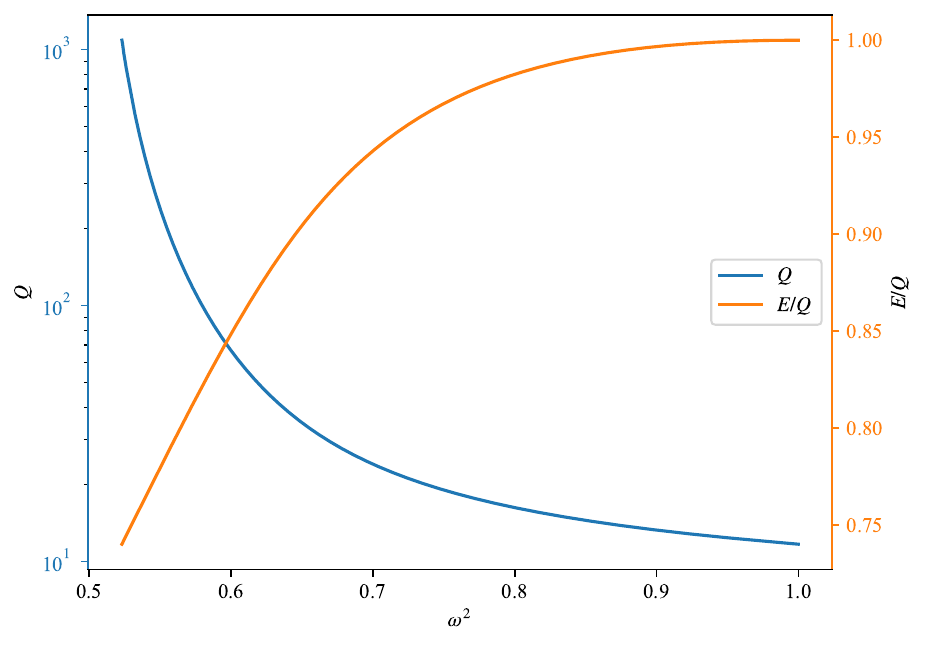}\label{fig:3}}
		\caption{Upper panel: Configurations of Q-strings with different oscillation frequencies. Lower panel: The charge and the radio of energy to charge as a function of the square of the oscillation frequency.}
		\label{fig:2-3}
	\end{center}
\end{figure}

To mimic a cylindrical fluid, we adopt the usual cylindrical coordinates $(t,\rho,\varphi,z)$ and a $z$-invariant ansatz without angular excitation for the scalar field $\psi=\phi(\rho)e^{-i\omega t}$,
where the profile function is determined by the following ordinary differential equation
\begin{equation}
	0=\frac{d^{2}\phi}{d\rho^{2}} + \frac{1}{\rho}\frac{d\phi}{d\rho} + \left(\omega^{2}  - V'\right) \phi,
\end{equation}
with $V'= \frac{dV}{d|\psi|^{2}}(\phi)$.
In the framework of Newtonian mechanics, such an equation describes the motion of a classical particle of unit mass, with position $\phi$ and time $\rho$, in the effective potential $U=\frac{1}{2}\left(\omega^{2}\phi^{2}-V\right)$,
and under the influence of the friction $F=-\frac{1}{\rho}\frac{d\phi}{d\rho}$.
In this consideration, a solution corresponds to a trajectory that starts from position $\phi=\phi_{0}$ at time $\rho=0$ and terminates at origin $\phi=0$ after infinite time $\rho\rightarrow\infty$, as shown in Figure \ref{fig:2}.
The existence of the trajectory requires that the mechanical energy of the particle at the initial position is greater than that at the end point. 
Combined with the requirement of the physical bound state, the oscillation frequency must be within the interval $\omega^{2}\in(0.5,1)$, where the upper and lower bounds correspond to the thick-wall and thin-wall limits respectively.

Figure \ref{fig:3} shows how the energy and charge per unit length, defined as follows
\begin{equation}
	\begin{aligned}
		E=&2\pi\int_{0}^{\infty} \rho \left[\omega^{2}\phi^{2}+\left(\frac{d\phi}{d\rho}\right)^{2}+V\right]d\rho,\\
		Q=&4\pi\omega\int_{0}^{\infty} \rho\phi^{2} d\rho,
	\end{aligned}
\end{equation}
relate to the oscillation frequency, from which two conclusions can be drawn.
(1) The monotonically decreasing charge indicates that the well-known stability criterion for solitons $dQ/d\omega<0$ is satisfied \cite{Lee:1991ax}.
Therefore, Q-strings are expected to be stable against fluctuations.
(2) The energy of a Q-string is always less than that of a collection of free particle quanta with the same particle number, namely $E<mQ$, indicating that it is prevented from decaying into free particles.
In this case, solitons are usually said to be absolutely stable \cite{Lee:1991ax}.

{\it Hydrodynamic instability.}---From the linear perturbation theory, the dynamical stability of the system is dominated by the linearized Klein-Gordon equation
\begin{equation}
	\left[-\partial^{2}_{t}+\Delta-V'-\phi^{2}V''\right]\delta\psi-e^{-2i\omega t}\phi^{2}V''\delta\psi^{*}=0, 
\end{equation}
where $\Delta$ is the three-dimensional Laplace operator.
Due to the existence of self-interaction, the perturbation $\delta\psi$ is coupled with its conjugate $\delta\psi^{*}$, resulting in the monochromatic wave failing to solve the above perturbation equation.
The correct ansatz for the perturbation should contain at least one pair of dichromatic waves,
\begin{equation}
	\delta\psi=e^{-i\omega_{+} t-ikz}\delta \psi_{+}\left(\rho\right) +e^{-i\omega_{-}^{*}t+ikz} \delta \psi_{-}\left(\rho\right),\label{eq:5}
\end{equation}
with frequencies $\omega_{\pm}=\omega\pm\Omega$.
The above ansatz involves the perturbation along the cylindrical direction with the wavelength $\lambda=2\pi/k$.
With the bound state boundary condition $\delta\psi_{\pm}(\rho\rightarrow\infty)=0$, the eigenfrequency is complex $\Omega=\Omega_{R}+i\Omega_{I}$.
The mode with a positive imaginary part $\Omega_{I}>0$ is dynamically unstable and grows exponentially in the form of $e^{\Omega_{I}t}$.

\begin{figure}
	\begin{center}
		\subfigure[]{\includegraphics[width=\linewidth]{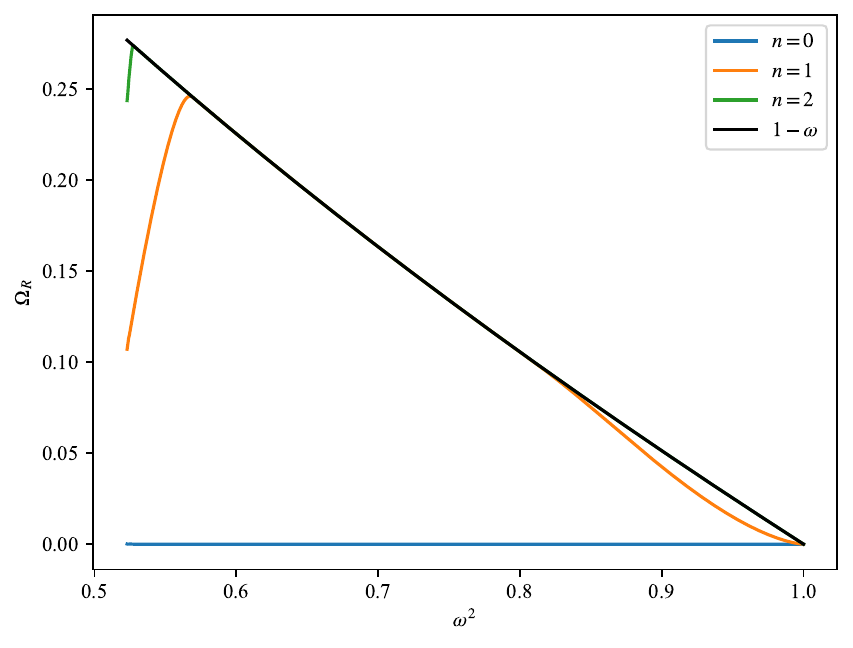}\label{fig:4}}
		\subfigure[]{\includegraphics[width=\linewidth]{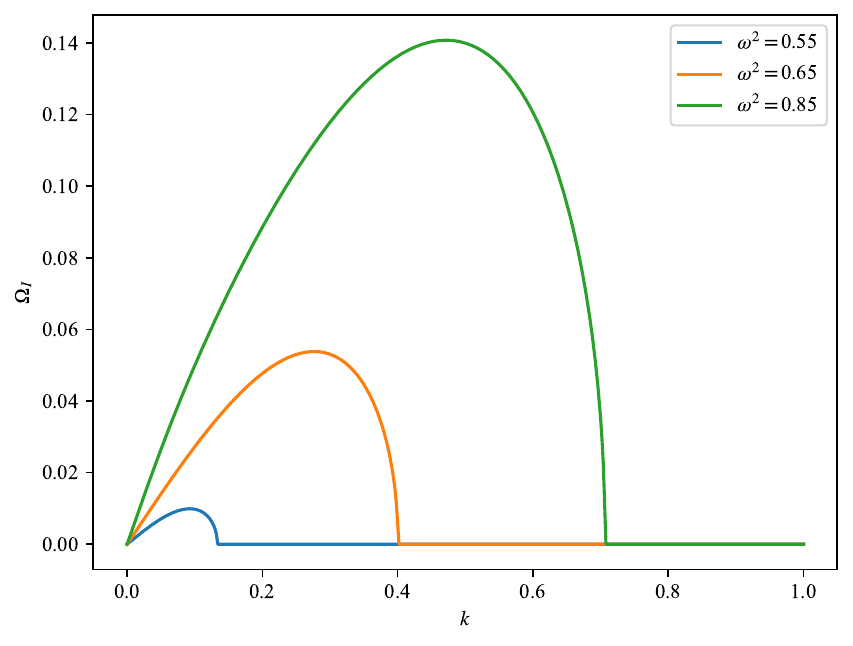}\label{fig:5}}
		\caption{Upper panel: The discrete spectrum of oscillation modes $n=1,2$ and a zero mode $n=0$ at $k=0$. Lower panel: The dispersion relation for unstable hydrodynamic modes. Curves of different colors represent Q-strings with different oscillation frequencies.}
		\label{fig:4-5}
	\end{center}
\end{figure}

In the case of $k=0$, the numerical results using the spectral decomposition method \cite{Boyd:2001} show that Q-strings are dynamically stable, consistent with the stability criterion for solitons.
The eigenfrequency $\Omega$ is discretely distributed within the interval $[0,1-\omega)$ along the real axis.
Figure \ref{fig:4} shows the first two oscillation modes, $n=1,2$.
The value of the integer $n$ corresponds to the number of nodes in the radial profile of the eigenstate.
The situation is similar to that of Q-balls \cite{Kovtun:2018jae}.
The number of oscillation modes increases within the thin-wall region.
Approaching the thick-wall limit, an isolated oscillation mode breaks in from outside the interval boundary, similar to what is observed near the cusp point in the case of Q-balls.
There is no oscillation mode in the intermediate region.
All Q-strings possess a zero mode $n=0$, which is a hydrodynamic mode, defined as $\Omega(k\rightarrow0)=0$.

These hydrodynamic modes are dynamically unstable to long-wavelength perturbations.
The dispersion relation is illustrated in Figure \ref{fig:5}.
Within the range of small $k$, the hydrodynamic mode transforms into a purely imaginary mode with dynamical instability $\Omega_{I}>0$, whose imaginary part grows linearly with the wave number.
The imaginary part of the unstable mode is called the growth rate, which reaches saturation at the optimal wave number.
After that, the instability is suppressed until it disappears. 
The unstable hydrodynamic mode turns into an oscillation mode.
We have checked that all oscillation modes are dynamically stable to perturbations of form (\ref{eq:5}).

This long-wavelength instability induced by hydrodynamic modes is similar to the sound mode instability of a viscous fluid with thermodynamic instability \cite{Janik:2015iry}.
In the long-wavelength region, the dispersion relation for a sound mode is $\Omega=\pm\sqrt{dp/d\epsilon}k-i\Gamma k^{2}+o(k^{3})$ with pressure $p$ and energy density $\epsilon$ \cite{Romatschke:2009im}.
The sound attenuation constant $\Gamma$ associated with viscosities is positive.
Therefore, the thermodynamic instability with $dp/d\epsilon<0$ will induce a dynamically unstable branch with a growth rate
\begin{equation}
	\Omega=\sqrt{|dp/d\epsilon|}ik+o(k^{2}).\label{eq:6}
\end{equation}
As the wave number increases, the quadratic term induced by viscosities gradually plays a dominant role, suppressing this instability.
For a Q-string, the central region is a uniform perfect fluid, which is thermodynamically stable $dp/d\epsilon>0$ in the case of a ground state with low energy density, as presented in the Appendix, indicating that the long-wavelength instability suffered by Q-strings is distinct from that related to thermodynamics.
We will next demonstrate that this instability is a membrane instability associated with the Rayleigh-Plateau phenomenon induced by interface effects, resulting in dynamical transitions in geometric configurations rather than in thermodynamic properties.
 
Similar to the Affleck-Dine mechanism, this instability provides a dynamical mechanism for the generation of Q-balls.
The exponential growth of perturbations with super-threshold wavelengths will decompose the Q-string into multiple separate Q-balls, similar to the formation process of droplets on a jet.
The size of final Q-balls is set by the mode with the largest imaginary part at the optimal wavelength, as it grows fastest and will soon dominate the dynamical process.
This scenario, inferred from fluid dynamics, needs to be further supported by dynamical simulations.

{\it Rayleigh-Plateau instability.}---For a jet of liquid without viscosity, the dispersion relation for the growth rate of the unstable Rayleigh-Plateau mode is as follows
\begin{equation}
	\Omega^{2}_{RP}=\frac{T}{\rho_{0}R^{3}}\frac{ikRJ'_{0}\left(ikR\right)}{J_{0}\left(ikR\right)}\left(1-k^{2}R^{2}\right),\label{eq:7}
\end{equation}
where $J$ is a Bessel function, $T$ and $\rho_{0}$ are the surface tension and density of the liquid respectively.
The threshold is equal to the inverse of the cylinder radius $k_{RP}=1/R$.
To compare with the above hydrodynamic instability, we define the surface tension and effective radius of Q-strings as follows \cite{Mai:2012yc}
\begin{equation}
	T=\int^{\infty}_{0}s(\rho)d\rho, \quad R=\frac{1}{T}\int^{\infty}_{0}\rho s(\rho)d\rho,
\end{equation}
with the shear forces $s(\rho)=2\phi'^{2}$.
Since the interior of Q-strings in the thin-wall region is almost uniform, we use the charge density at the center to replace the liquid density $\rho_{0}$ in (\ref{eq:7}).

\begin{figure}
	\begin{center}
		\subfigure[]{\includegraphics[width=\linewidth]{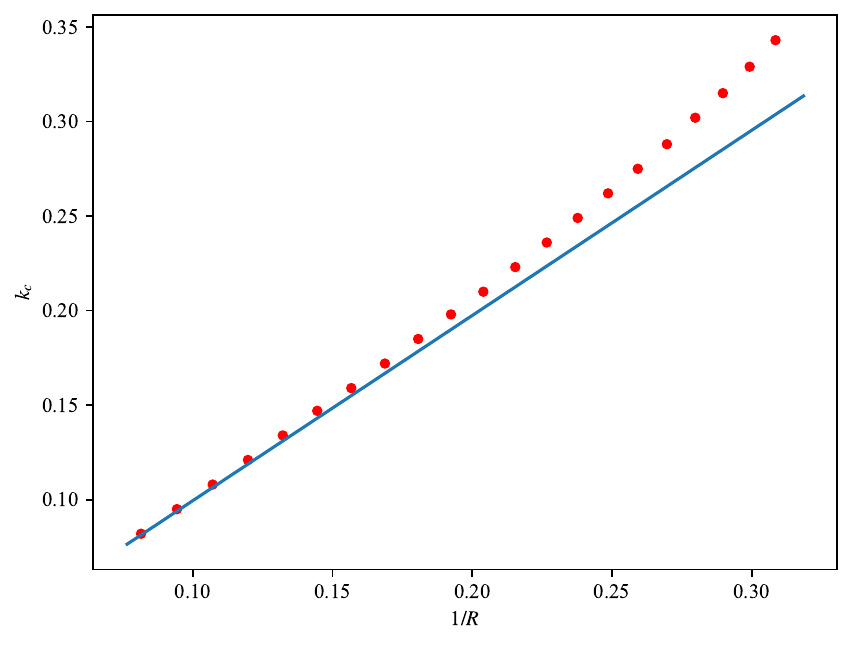}\label{fig:6}}
		\subfigure[]{\includegraphics[width=\linewidth]{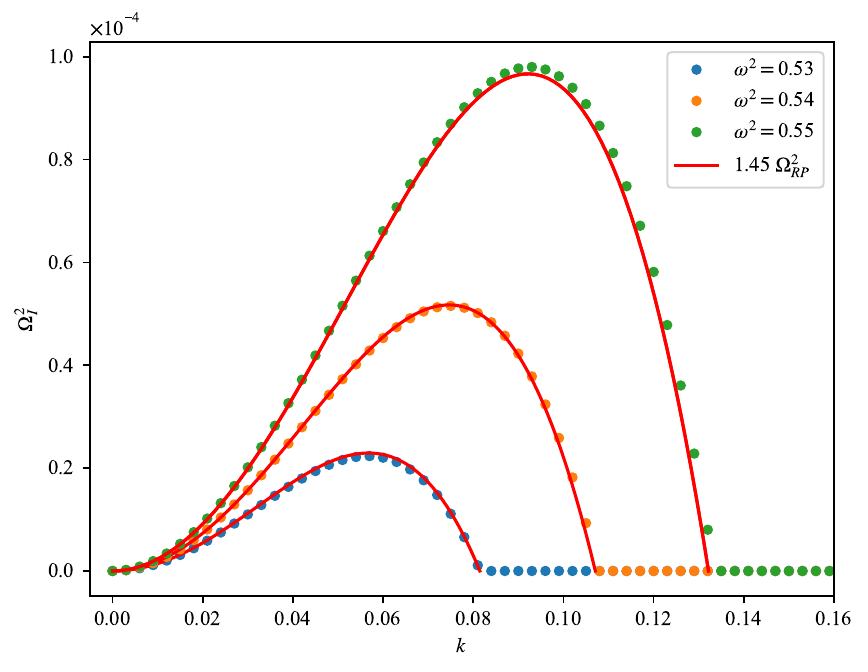}\label{fig:7}}
		\caption{Upper panel: The threshold $k_{c}$ of hydrodynamic instability as a function of the inverse of the radius $1/R$ for Q-strings in the thin-wall region. The blue line indicates $k_{c}=1/R$. Lower panel: The dispersion relations of the hydrodynamic instability and the Rayleigh-Plateau instability. Dots of different colors represent Q-strings with different oscillation frequencies.}
		\label{fig:6-7}
	\end{center}
\end{figure}

In the thin-wall region, the interface of Q-strings approaches a membrane, as shown in \ref{fig:2}, matching the cylindrical fluid jet in the Rayleigh-Plateau model.
In this case, the threshold of the hydrodynamic instability suffered by Q-strings approaches that of the Rayleigh-Plateau instability, equal to the reciprocal of the geometric radius, as shown in Figure \ref{fig:6}.
Away from the thin-wall region, the Rayleigh-Plateau analogy without considering the thickness of the interface is no longer applicable, resulting in the deviation in the figure.
Not only the thresholds, the dispersion relations of these two instabilities are in perfect agreement, as shown in Figure \ref{fig:7}, except for an overall factor of $1.45$, which may arise from the properties of solitons.
Such duality reveals the hydrodynamic properties of solitons, at least near equilibrium.
Since the effect of viscosity is not considered in the dispersion relation (\ref{eq:7}), the interface of the Q-string is similar to a low-viscosity fluid, like a black string, providing a theoretical basis for the preparation of Q-objects in superfluids \cite{Bunkov:2007fe}.


{\it Discussions.}---The similarities between Q-objects and gravitational black objects pose challenges for astronomical observations.
Black holes aside, replacing the supermassive compact object at the center of the Milky Way with a Q-ball does not seem to be catastrophic \cite{Troitsky:2015mda}.
An energy extraction mechanism similar to black hole superradiance also occurs in Q-objects \cite{Saffin:2022tub,Cardoso:2023dtm}.
Unfortunately, our results show that the interface of Q-objects is similar to the event horizon of black objects, with both exhibiting low-viscosity hydrodynamic behavior.
One possible distinction is that the interface of solitons are elastic and will bounce physical signals, causing echoes \cite{Cardoso:2019rvt}, rather than absorbing everything like the event horizon.

Unlike black strings, whose instability occurs in 5 and higher dimensions, 4-dimensional Q-strings exhibit a long-wavelength instability that is of more experimental interest.
There are various ways to generate string-like Q-objects, such as the Affleck-Dine mechanism \cite{Enqvist:1999mv,Kasuya:1999wu,Kasuya:2000wx} with directional perturbations, the dynamics \cite{Axenides:1999hs,Battye:2000qj} and charge-swapping behavior \cite{Copeland:2014qra} of solitons.
Our findings theoretically explain the metastability of solitons with local string configurations, which will exhibit two evolution patterns: shrinking or splitting, depending on whether the local length is larger than the instability threshold.
A typical example is the Q-ring, which will further collapse or split \cite{Axenides:2001pi}.
Another more general example is the filamentary structure in the Affleck-Dine condensation, which can fragment to form Q-balls \cite{Enqvist:2000cq,Hiramatsu:2010dx}.
These string-like Q-objects, serving as dynamical intermediate states, will reset the size of Q-balls.

There are still many aspects that remain open.
One direction is to further verify whether the dynamical characteristics of the interface of Q-objects away from equilibrium can be fully described by second-order hydrodynamics of viscous fluids.
If so, in turn, the soliton model is an alternative to the hydrodynamic framework, providing a hypothetical field theory for fluid motion.
On the other hand, in the effective worldvolume theory \cite{Emparan:2009cs,Emparan:2009at}, the Gregory-Laflamme instability is associated with a local thermodynamic instability, leaving the question of whether there is a thermodynamic framework similar to that of black holes for solitons that reveals the thermodynamic duality of the hydrodynamic instability suffered by Q-strings.

\noindent{\bf Acknowledgments.}---The author thanks Lars Andersson for helpful discussions and acknowledges the support from the National Natural Science Foundation of China with Grant No. 12447129 and the fellowship from the China Postdoctoral Science Foundation with Grant No. 2024M760691.

\noindent The data that support the findings of this article are openly available \cite{Chen:data}.

\noindent{\bf Appendix: soliton/fluid duality.}---The hydrodynamic instability of Q-strings provides a field theory explanation for the Rayleigh-Plateau phenomenon of classical fluids, indicating a duality framework between solitonic field theory and fluid dynamics.
The purpose of this appendix is to present more specialized details for such a duality framework.

In fluids, there are two types of dynamical instabilities in the long-wavelength region, leading to the violation of spatial symmetry.
One is related to the intrinsic thermodynamic properties of the fluid: the sound mode instability driven by the thermodynamic instability $dp/d\epsilon<0$, with a growth rate (\ref{eq:6}).
The other is derived from the interface effects of the fluid: the membrane instability driven by surface tension, with the Rayleigh-Plateau dispersion relation (\ref{eq:7}).
The difference in the internal mechanism results in different physical phenomena.
The former triggers a dynamical transition in the thermodynamic properties of the fluid, manifested as the generation of a new thermal phase with the thermodynamic stability $dp/d\epsilon>0$.
This dynamical process usually leads to the generation of interfaces, the shape of which depends on the geometry of the perturbations, serving as seeds for condensation, such as the holographic fluid in AdS/CFT theory \cite{Janik:2015iry,Janik:2017ykj}.
The latter induces a dynamical transition in the geometric configuration of the fluid interface, from cylindrical to spherical.
This is why the end of a water column coming out of a tap breaks up into droplets, and is responsible for the instability of black strings in general relativity with dimensions $D\geqslant 5$ \cite{Gregory:1993vy,Cardoso:2006ks,Lehner:2010pn}.

Interestingly, these two hydrodynamic instabilities correspond exactly to the two generation mechanisms of Q-balls with spherical interfaces in the solitonic field theory.
Consider the Affleck-Dine mechanism, a coherently oscillating scalar condensate dominated by a potential that is shallower than the quadratic self-interaction $U(\phi)\propto\phi^{2+2K}$ with $K<0$ and $|K|<1$ \cite{Enqvist:1997si}, with the equation of state 
\begin{equation}
	p=\frac{K}{2+K}\epsilon.\label{eq:9}
\end{equation}
From the linear analysis, the negative parameter $K$, indicating the thermodynamic instability $dp/d\epsilon<0$, induces a dynamical instability with long-wavelength in such a scalar condensate, with a growth rate
\begin{equation}
	\Omega = \sqrt{\frac{|K|}{2+K}}ik+o(k^{2}),
\end{equation}
which, combined with the relation (\ref{eq:9}), is completely consistent with the sound mode instability  (\ref{eq:6}) in fluids.
As a result, a new thermal phase, Q-matter, is dynamically generated from the Affleck-Dine field \cite{Enqvist:1999mv,Kasuya:1999wu,Kasuya:2000wx}.

\begin{figure}
		\includegraphics[width=\linewidth]{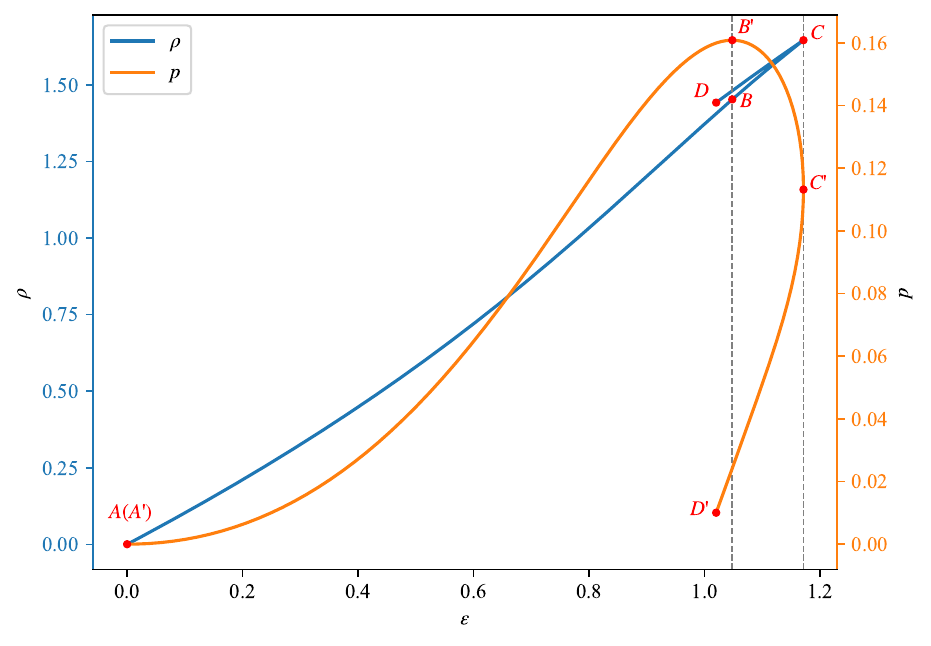}
		\caption{In the central region of Q-strings, the charge density and pressure as a function of the energy density. Points $A(A')$ and $D(D')$ represent the thick-wall limit and thin-wall limit respectively.}
		\label{fig:8}
\end{figure}

To be immune to this instability, the resulting Q-matter should be a thermal phase with the thermodynamic stability $dp/d\epsilon>0$.
For the counterpart of the cylindrical interface, the Q-string, the central region is a uniform perfect fluid with pressure $p=\omega^{2}\phi^{2}_{0}-V(\phi_{0})$, energy density $\epsilon=\omega^{2}\phi^{2}_{0}+V(\phi_{0})$, and charge density $\rho=2\omega\phi^{2}_{0}$.
Figure \ref{fig:8} shows the relationship between these physical quantities, in which there are three regions: thick-wall region $AB$, middle region $BC$ and thin-wall region $CD$.
Among them, the middle region has the same charge density as the thin-wall region, but a greater energy density.
Since the system tends to reside in a ground state with lower energy, the formed Q-strings should be located in the thick-wall and thin-wall regions, both of which are thermodynamically stable $dp/d\epsilon>0$, as expected.
Combined with the results presented in this letter, one can conclude that the hydrodynamic instability suffered by Q-strings is a membrane instability associated with the Rayleigh-Plateau phenomenon, originating from interface effects and independent of internal properties.
Expectedly, the interface of Q-strings will deform as the instability grows, leading to the formation of Q-balls, similar to the growth of droplets in a liquid jet.

The above signs strongly imply the existence of a duality relationship between solitonic field theory and fluid dynamics, facilitating further revelation of the commonalities between the two fields.
Benefiting from this, researchers in fluids can approximately capture the dynamical behavior of fluids through soliton models, and researchers in solitons can study the transport properties of solitons in non-equilibrium processes through a hydrodynamic framework.

\end{document}